\documentstyle[12pt]{article}

\def\be{\begin{equation}}
\def\ee{\end{equation}}
\def\br{\begin{eqnarray}}
\def\er{\end{eqnarray}}
\def\nonu{\nonumber}
\def\br{\begin{eqnarray}}
\def\er{\end{eqnarray}}
\def\vp{\varphi}

\def\RR{{\rm I\kern-.1567em R}}                              
 \def\CC{{\rm C\kern-4.7pt                                    
 \vrule height 7.7pt width 0.4pt depth -0.5pt \phantom {.}}} 
 \def\ZZ{{\sf Z\kern-4.5pt Z}}                                

\begin{document}

\begin{titlepage}
\vspace*{-2 cm}
\noindent

\vskip 3cm
\begin{center}
{\Large\bf Hopf maps as static solutions of the complex eikonal 
equation }
\vglue 1  true cm
C. Adam 

\vspace{1 cm}
 {\footnotesize Departamento de F\'\i sica de Part\'\i culas,\\
Facultad de F\'\i sica\\
Universidad de Santiago\\
E-15706 Santiago de Compostela, Spain}

\medskip
\end{center}

\normalsize
\vskip 0.2cm

\begin{abstract}
We demonstrate that a class of torus-shaped Hopf maps with arbitrary linking
number obeys the static complex eikonal equation. Further, we explore the
geometric structure behind these solutions, explaining thereby the
reason for their existence. As this equation shows
up as an integrability condition in certain non-linear
field theories, the existence of such solutions is of some interest. 
\end{abstract}

\vskip 2cm

\end{titlepage}

\section{Introduction}
In this article we want to report on a class of Hopf maps with arbitrary
linking number, which are, at the same time, static solutions to the
complex eikonal equation. Further, we want to explore the geometric
structure which is behind these solutions and explains, in fact, 
their existence.

The eikonal equation reads
\be \label{eik}
(\partial^\mu \chi )(\partial_\mu \chi ) =0
\ee
and describes, for a real scalar function $\chi$, 
the propagation of wave fronts
(field discontinuities) in Minkowski space. 
Its generalization to complex $\chi$ has some applications in
 optics and quantum mechanics, as well as in general relativity 
(see \cite{Kass} and the literature cited there).
In \cite{Kass} an algebraic procedure (based on twistor methods) 
for the construction of complex solutions
to Eq. (\ref{eik}) was developed, and some examples of singular solutions
were provided.
The complex eikonal equation admits even static solutions, i.e., 
solutions to the equation
\be \label{stat-eik}
(\nabla \chi )\cdot (\nabla \chi ) =0,
\ee
in contrast to the case of real $\chi$. 

The complex eikonal equation (\ref{eik}) has also appeared, in quite a
different context,  as an integrability
condition in some non-linear field theories. 
In the last few years there has been rising interest in integrable field 
theories in higher (i.e., more than two) dimensions, where
solitonic solutions are, in many cases, provided by 
certain Hopf maps, see e.g.
\cite{noso,afz,baf,asg}. In addition, some non-linear field theories which are,
in general, not integrable, contain {\em integrable subsectors}
where certain integrability conditions are satisfied.
Specifically, the complex eikonal equation (\ref{eik}) defines
integrable subsectors in the Skyrme and Skyrme--Faddeev models
(\cite{FSG1,FSG2}). For static, solitonic solutions, this condition 
reduces to the
static complex eikonal equation (\ref{stat-eik}). Finite energy solitons 
in these integrable subsectors correspond
to static solutions defined on one-point compactified $\RR^3$ and
may, therefore, be identified with functions on $S^3$ via
stereographic projection. Further,  the target space 
 of the fields $\chi$ in the integrable subsectors
can be identified with the Riemann sphere $S^2$ (i.e., $\chi$ is
a holomorphic variable on $\CC$). Therefore,  the fields $\chi$ in the
integrable subsectors of these models are
Hopf maps $S^3 \to S^2$, and can be classified by the Hopf index (the
homotopy group $\pi_3 (S^2)=\ZZ$). Consequently, solutions of the 
complex, static eikonal equation which are, at the same time, Hopf maps, are
of some interest for these non-linear field theories, because they
provide finite energy field configurations in their integrable subsectors.

In addition, the static complex eikonal equation (\ref{stat-eik}) has
appeared as an integrability condition for the existence of multiple
zero modes of the static, Abelian Dirac operator (\cite{AMN1,AMN2,ES1}).
In this case,  solutions $\chi$ to Eq. (\ref{stat-eik}) are again 
required to be
Hopf maps. The Hopf maps described below will indeed 
give rise to the construction of new classes of zero modes with new
and interesting properties, but this issue shall be discussed elsewhere.

In Section 2 we show that certain toroidal Hopf maps $\chi^{(m,n)}$ obey the
static eikonal equation for arbitrary integer $m$ and $n$
(here $m$ and $n$ count the number of times the level curves of $\chi$ 
wrap around the two circular directions of a certain torus). 
Further, we briefly discuss the symmetries of the static eikonal equation,
which enables us to construct new solutions from the ones just mentioned.

In Section 3 we explain the geometric structure which lies behind the
existence of these solutions. It turns out that the solutions of
Section 2 may be understood as pullbacks of trivial
solutions of the complex eikonal equation in two dimensions which 
preserve some metric properties, providing thereby non-trivial
three-dimensional solutions. Further, we give a sufficient condition for
the existence of solutions of the geometric type discussed in this paper.

\section{The solutions}

In the sequel, we will express Hopf maps as complex-valued functions
which depend on three variables like, e.g., $(x,y,z)$. Here, the space
spanned by these variables may be interpreted either as one-point
compactified $\RR^3$ or as the three-sphere $S^3$, where a stereographic
projection has been performed. The solutions to the static eikonal
equation described below do not depend on this interpretation, i.e.,
they may be interpreted as solutions on $\RR^3$ or on $S^3$. This result 
is related to the fact that the metrics on $\RR^3$ and $S^3$ are
conformally equivalent (i.e., equal up to a local, space-dependent
scale transformation),
as will become clear in the next section.    

The simplest Hopf map is
\be \label{s-ho}
\chi^{(1,1)} =i\frac{2(x +iy)}{2z +i(r^2 -1)} 
\ee
(the meaning of the superscript $(1,1)$ 
is explained below in Eq. (\ref{chimn})). Further,
$r^2 \equiv x^2 +y^2 +z^2$, and the irrelevant pre-factor $i$ has 
been chosen for later convenience. The simplest Hopf map
is well-known to obey the static eikonal equation (\ref{stat-eik}), see
e.g. \cite{AMN1}. Before demonstrating this fact, we want to
introduce toroidal coordinates $(\eta ,\xi ,\varphi)$ via
\br
x &=&  q^{-1} \sinh \eta \cos \vp \;\;, \;\;
y =  q^{-1} \sinh \eta \sin \vp   \nonu \\
z &=&  q^{-1} \sin \xi \quad ;  \qquad  q = \cosh \eta - \cos \xi .
\label{tordefs}
\er
Further, we need the gradient in terms of the toroidal coordinates,
\be \label{grad-3}
\nabla = (\nabla \eta)\partial_\eta 
+(\nabla \xi )\partial_\xi +(\nabla \varphi)\partial_\varphi
= q(\hat e_\eta \partial_\eta + \hat e_\xi \partial_\xi +
\frac{1}{\sinh \eta} \hat
e_\varphi \partial_\varphi  )
\ee
where $(\hat e_\eta  ,\hat e_\xi ,\hat e_\varphi )$ form an orthonormal
frame in $\RR^3$.  
In terms of toroidal coordinates, the simplest Hopf map  reads
\be \label{chi11}
\chi^{(1,1)} =\sinh \eta \, e^{i\varphi +i\xi }.
\ee        
Here, surfaces of $\eta = {\rm const.}$ are tori in $\RR^3$. These tori
are rotation symmetric around the $z$ axis, and all of them enclose
the circle $ C=\{\vec x \in \RR^3 :\;
  z=0 \;
\wedge \; r^2 =1\}$. The coordinates $\varphi$ and $\xi$ are angular
coordinates along the two circular directions on each torus.
Each level curve of $\chi^{(1,1)}$ (i.e., each curve 
$\chi^{(1,1)}={\rm const.}$) is
located on one torus. It is, in fact, a
circle that winds once around each circular direction of the torus.
Further, any two different level curves are linked with linking number
one, and this linking number is the geometric definition of the
Hopf index (which is equal to one for the simplest Hopf map (\ref{s-ho})).    

For a simple demonstration of the fact that the Hopf map (\ref{chi11}) 
really obeys the eikonal equation it is useful to 
re-express a general Hopf map $\chi$ in terms of two real
functions (modulus $S$ and phase $\sigma$) like
\be
\chi = S\, e^{i\sigma} .
\ee
In terms of these real functions, the static eikonal equation 
(\ref{stat-eik}) leads to the conditions
\be \label{re-eik}
(\nabla S)\cdot (\nabla \sigma )=0 \, ,\qquad (\nabla S)^2 =S^2
(\nabla \sigma )^2 .
\ee
For the simplest Hopf map (\ref{chi11}) we find, with $S=\sinh \eta$, $\sigma 
=\xi +\varphi$,
\be \label{eik-chi11}
\nabla S = q \cosh \eta \, \hat e_\eta \, ,\quad 
\nabla \sigma = q\left( \hat e_\xi + \frac{1}{\sinh \eta}
\hat e_\varphi \right)
\ee
which indeed obey Eqs. (\ref{re-eik}). The important point here is that 
the equations (\ref{re-eik}) are expressed only in terms of the target
space coordinates $S$ and $\sigma$, making the problem essentially 
two-dimensional. This is precisely what happens for the simplest Hopf map.
The factor $q$, which is present in (\ref{eik-chi11}) and cannot be
expressed in terms of the target space coordinates, cancels in the relations
(\ref{re-eik}).
 
A simple generalization to higher Hopf maps is provided by the functions
\be \label{chimn}
\chi^{(m,n)} = f(\eta)\, e^{ in\varphi + im\xi} \quad , \quad m,n \in \ZZ
\ee
which are true Hopf maps if the real function $f$ obeys certain regularity
conditions like, e.g., $f(0)=0$ and $f(\infty )=\infty$ (what we assume
in the sequel). The level curves of these Hopf maps still lie on the same
tori as above, but now they wind $n$ times around the $\varphi$ 
direction and $m$ times around the $\xi$ direction. Further, the Hopf
index $N_{\rm H}$
(i.e., the linking number of any two different level curves) is
$N_{\rm H} =nm$. 

We find for the gradient
\be
\nabla \chi^{(m,n)} = q\, e^{im\xi +in\varphi} 
\left( f' \hat e_\eta +imf\hat e_\xi +
\frac{in}{\sinh \eta} f\hat e_\varphi \right) ,
\ee
where $f' \equiv \partial_\eta f$, 
and Eq. (\ref{stat-eik}) leads to the simple differential equation
\be \label{eq-efet}
\frac{f'}{f}= \left( m^2 +\frac{n^2}{
\sinh^2 \eta }\right)^\frac{1}{2}
\ee 
with the solution
\be \label{hi-sol}
f=\sinh^{|n|} \eta \, \, 
\frac{\left( |m|\cosh \eta + \sqrt{n^2 +m^2 \sinh^2 \eta
}\right)^{|m|}}{\left( |n|\cosh \eta + \sqrt{n^2 +m^2 \sinh^2 \eta
}\right)^{|n|}} \, .
\ee
These solutions are genuine Hopf maps for all non-zero, integer $m,n$, 
because $f$ obeys $f(0)=0$, $f(\infty) =\infty$. 

At this point it is of interest to briefly consider the symmetries of
the complex static eikonal equation (\ref{stat-eik}). This will lead to 
some further understanding of these solutions and allow to construct more
solutions from the ones obtained so far. The symmetry group of equation
(\ref{stat-eik}) is a direct product of base space and target space symmetries,
where the group of base space symmetries is the conformal group in 
three-dimensional Euclidean space. The group of target space symmetries
is given locally by the maps $\chi \rightarrow F(\chi)$, 
where $F$ is an arbitrary
complex function of $\chi$, but not of its complex conjugate $\bar \chi$. 
The requirement that the solutions $\chi' = F(\chi)$ are single-valued
again restricts the allowed functions $F(\cdot )$ to the set of
holomorphic functions on $\CC$.

The presence of the conformal symmetry on base space implies that the
ansatz (\ref{chimn}) is an ``educated guess'' for a solution to Eq.
(\ref{stat-eik}) in the sense of the Lie theory of symmetry. That is
to say, if we choose a rotation about the $z$ axis and a certain
combination of proper conformal transformation along the $z$ axis and 
translation along the $z$ axis as a maximal set of two commuting base 
space transformations, then the corresponding infinitesimal 
symmetry generators
(vectors ${\bf v}^i$) are precisely given by the tangent vectors along
$\varphi$ and $\xi$, ${\bf v}^1 =\partial_\varphi$, and ${\bf v}^2 =
\partial_\xi $. The ansatz (\ref{chimn}) is invariant under a combination
of these base space transformations and  phase transformations of the 
target space variable $\chi$, i.e., under the action of the vector
fields $\tilde {\bf v}^1 = \partial_\varphi -in\chi \partial_\chi$ and
$\tilde {\bf v}^2 = \partial_\xi -im\chi \partial_\chi$, which provides
precisely the educated guess according to Lie. A concise discussion of
these points can be found in Ref. 4, where the symmetries of an integrable
model with infinitely many Hopf solitons are discussed in detail.

Further, we may use the target space symmetries to construct more solutions
from the ones given in (\ref{hi-sol}). In fact, each field $\chi' 
=F(\chi)$ is a solution, where $\chi$ is a solution and $F$ is a
holomorphic function on $\CC$.

\section{Geometric background}

Here we want to explain the geometric structure behind the solutions
(\ref{hi-sol}),
which will, in fact, allow to understand the reason why they exist. 
For this purpose, let us first 
observe that there exist trivial solutions to the
complex eikonal equation in $\RR^2$ or, equivalently, in $\CC$. 
Indeed, for real, cartesian coordinates $(u,v)\in \RR^2$ with 
$w=u+iv$ and gradient
\be \label{grad-2}
\nabla^{(2)}\equiv \hat e_u \partial_u +\hat e_v \partial_v
\ee
the complex eikonal equation $(\nabla^{(2)} f(w))^2 =0$ is equivalent to
the Cauchy-Riemann equations, which are obeyed by arbitrary
holomorphic functions $f(w)$.
So, obviously, the complex coordinate $w=u+iv$ itself obeys the
eikonal equation,
\be
(\nabla^{(2)} w)^2 =0.
\ee
By introducing the modulus $\rho$ and phase $\phi$ of $w$, 
\be \label{mo-ph-2}
w=\rho e^{i\phi} ,
\ee
this equation
leads to the conditions
\be \label{re2-eik}
(\nabla^{(2)} \rho )\cdot (\nabla^{(2)} \phi )=0 \, ,\qquad
(\nabla^{(2)} \rho)^2 =\rho^2 (\nabla^{(2)} \phi )^2 .
\ee
It holds in fact also that
\be
(\nabla^{(2)} \rho)^2 =\rho^2 (\nabla^{(2)} \phi )^2 =1.
\ee
Conditions (\ref{re2-eik}) 
are completely analogous to the conditions (\ref{re-eik}) in 
three dimensions. This leads to the natural assumption that the
conditions (\ref{re-eik}) in three dimensions are just the pullbacks
under the Hopf map $\chi$ of the two-dimensional conditions
(\ref{re2-eik}). In the sequel we want to show that this is true in
a specific sense.

For this purpose, we want to re-express the above remarks in a more 
geometric fashion, where we introduce the metrics of the spaces under
consideration and replace the gradients by exterior derivatives.

The metric 
on  the space $\RR^2$ is
\be \label{met2d}
g^{(2)}= d\rho \otimes d\rho + \rho^2 d\phi \otimes d\phi
\ee
and the dual metric is
\be
G^{(2)} = \partial_\rho \otimes \partial_\rho +\frac{1}{\rho^2}
\partial_\phi \otimes \partial_\phi .
\ee
The conditions (\ref{re2-eik}) translate into
\be
G^{(2)}(d\rho ,d\phi )=0 \, ,\quad G^{(2)}(d\rho ,d\rho )=
\rho^2 G^{(2)}(d\phi ,d\phi ) =1
\ee
and are obviously true. 

The metric in $\RR^3$ is
\br \label{met3d}
g &=& q^{-2}(d\eta \otimes d\eta + d\xi \otimes d\xi + \sinh^2 \eta
d\varphi \otimes d\varphi ) \nonumber \\
&=& \frac{q^{-2}}{1+t^2} [dt\otimes dt +(1+t^2) d\xi \otimes d\xi
+ t^2 (1+t^2) d\varphi \otimes d\varphi ]
\er
where $(\eta ,\xi ,\varphi)$ are the toroidal coordinates 
(see (\ref{tordefs})) and the coordinate 
\be
t=\sinh \eta
\ee
was introduced for later convenience. The dual metric is
\be
G=(1+t^2)q^2 \left( \partial_t\otimes \partial_t +\frac{1}{1+t^2} 
\partial_\xi \otimes \partial_\xi
+ \frac{1}{t^2 (1+t^2)} \partial_\varphi \otimes \partial_\varphi \right) .
\ee
As a next step we need the observation that a Hopf map $\chi$ introduces
a fiber-bundle structure on one-point compactified $\RR^3$ (or, equivalently,
on $S^3$). Here, the fibers are the level curves of the Hopf map. 
The fiber has the topology of the circle $S^1$, and the base space
has the topology of the sphere $S^2$ for all Hopf maps, but the induced
metric properties depend on the specific Hopf map.

Further, the Hopf map allows for a decomposition of the tangent bundle
$TM$ of
the fiber bundle 
$M= \RR^3$ (or $S^3$) into vertical and horizontal 
directions at each point of $M$.
Thereby two subbundles of the full tangent bundle $TM$ are induced,
which are called the vertical distribution $V$ and the horizontal 
distribution $H$.  
The vertical direction at each point
points along the fiber and is spanned (in our case) by
one vertor field $e_3$ which is pushed forward to zero under the Hopf map,
$\chi_* e_3 =0$.  The horizontal directions are spanned 
(in our case) by two vector fields $e_1$, $e_2$, 
which are perpendicular to the 
vertical vector $e_3$.
Obviously, the vertical direction only depends
on the Hopf map, whereas the horizontal directions depend on the
bundle metric, as well. Further, we will choose all three vectors $e_i$ to
have unit length (this condition depends, of course, on the metric).
This decomposition leads to an analogous decomposition at each point
$p\in M$ of the cotangent space $T_p^* M$ 
 into a vertical direction spanned by $\omega_3$ and horizontal
directions spanned by $\omega_1$ and $\omega_2$, where the $\omega_i$
are defined via 
\be
(\omega_i ,e_j) =\delta_{ij}
\ee
and $(\cdot ,\cdot )$ denotes the canonical inner product. 

Finally, the decomposition of the tangent space (and the cotangent space) 
into vertical and 
horizontal directions allows for a corresponding decomposition of the
metric and its dual into a vertical and a
horizontal component, $g=g_{\rm v} + g_{\rm h}$. They may be expressed
like
\be
g_{\rm h} = \omega_1 \otimes \omega_1 + \omega_2 \otimes \omega_2
\, ,\quad g_{\rm v}= \omega_3 \otimes \omega_3
\ee
\be
G_{\rm h} = e_1 \otimes e_1 + e_2 \otimes e_2 \, ,\quad G_{\rm v}
= e_3 \otimes e_3
\ee
in terms of the above vector fields and one-forms (observe that this 
notation just expresses the metric in terms of vielbeins in a 
coordinate-independent way).

Now we are in a position, eventually, to formulate sufficient conditions
for the existence of solutions to the conditions (\ref{re-eik}).

One sufficient condition is like follows: obviously, the push-forward
$\chi_*$ of the Hopf map defines an isomorphism from 
vectors in the horizontal distribution $H$ of
$TM$ at points $\vec x$ to vectors in $TN$ at points $\chi (\vec x)$
(here $N$ is the target space
manifold, i.e. $\CC$ or $S^2$, and $M$ is the fiber bundle). 
Now assume that this isomorphism is,
at the same time, an isometry, i.e., the lenght 
$| \chi_* v|$ of a pushed-forward vector field $\chi_* v$ 
in $TN$ w.r.t. the metric $g^{(2)}$ on $N$ at points $\chi (\vec x)$
is equal to the lenght
$| v|$ of an arbitrary horizontal vector field $v$ in $H$ with respect to the 
horizontal metric $g_{\rm h}$ at points $\vec x$.  Then, obviously, the
lengths of one-forms remain invariant under the pull-back $\chi^*$. For
a Hopf map $w=\chi (\vec x)$, which reads, in terms of real coordinates, like 
\be
\rho = S(\vec x) \, ,\quad \phi = \sigma (\vec x)
\ee
this means that the lenghts should pull back like
\be
|d\rho | = |dS| \, , \quad |d\phi | = |d\sigma |
\ee
and the target space metric $g^{(2)}$ expressed in coordinates $\rho ,\phi$
should be identical to the horizontal metric $g_{\rm h}$ expressed in
coordinates $S,\sigma $. Obviously, length relations are now conserved 
under the pull-back, as well,
\be \label{l-rel1}
|d \rho|^2 = \rho^2 |d\phi |^2 \quad \Rightarrow \quad
|dS|^2 = S^2 |d\sigma |^2 ,
\ee    
\be \label{l-rel2}
G^{(2)} (d\rho ,d\phi ) =0 \quad \Rightarrow G_{\rm h} (dS ,d \sigma )
=0,
\ee
which is precisely what we need in order to have solutions to the conditions 
(\ref{re-eik}). [Maps $\chi$ such that the push-forward $\chi_* \, :
\, H\to TN$ is an isometry are called Riemannian submersions and are described
at length e.g. in Ref. 11.]

It turns out that the condition on the Hopf map $\chi$ to be a Riemannian
submersion is too strong for our purposes. But there is a simple 
generalization which does just what we want. Suppose that the lengths
of horizontal vector fields are multiplied by a {\em common} factor at 
each point under
the push-forward, instead of being invariant. Then the lenghts of
one-forms will be multiplied by a common factor under the pull-back,
and this is sufficient for the conservation of the length relations
(\ref{l-rel1}), (\ref{l-rel2}) under the pull-back. For the
horizontal metric $g_{\rm h}$ and the target space metric $g^{(2)}$
this implies that they should be conformally equivalent, i.e., equal
up to a local scale factor. 
This is precisely what happens for our solutions, as
we want to demonstrate now explicitly.

First, we want to demonstrate it for the simplest Hopf map (\ref{chi11}).
We re-display the metric in $\RR^3$,
\be
g = [dt\otimes dt +(1+t^2) d\xi \otimes d\xi
+ t^2 (1+t^2) d\varphi \otimes d\varphi ] 
\ee
where we already ignored an irrelevant local scale factor, see (\ref{met3d}).
For the Hopf map $\chi =Se^{i\sigma}$ with $S=t$, $\sigma = \xi + \varphi$,
the vertical unit vector field $e_3$ is
\be
e_3 = \frac{1}{1+t^2} (\partial_\xi - \partial_\varphi )
\ee
(remember that $e_3 (\sigma ) = e_3 (S) =0$). 
The horizontal unit vector fields 
may be chosen as
\be
e_1 = \partial_t \, ,\quad e_2 = \frac{t}{1+t^2} (\partial_\xi + t^{-2}
\partial_\varphi ).
\ee
The corresponding vertical and horizontal one-forms are
\be
\omega_1 = dt \, ,\quad \omega_2 = t(d\xi + d\varphi )
\ee
\be
\omega_3 = d\xi - t^2  d\varphi 
\ee
and the horizontal metric is
\be
g_{\rm h} = \omega_1 \otimes \omega_1 + \omega_2 \otimes \omega_2 
= dt \otimes dt + t^2 (d\xi + d\varphi )\otimes (d\xi + d\varphi ) .
\ee
Obviously, this is identical to the target space metric (\ref{met2d})
once the identification $\rho \rightarrow t$, $\phi \rightarrow 
\sigma = \xi +\varphi $
is made.

Now we repeat this procedure for the class of Hopf maps $S=t$, $\sigma =
m\xi + n\varphi$ which are genuine Hopf maps with toroidal symmetry,
but not yet the solutions (\ref{hi-sol}).
We find for the horizontal and vertical unit vector fields
\be
e_1 = \partial_t \, ,\quad e_2 = \frac{t}{\sqrt{(n^2 +m^2 t^2 )(1+t^2)}}
(m\partial_\xi + nt^{-2}\partial_\varphi )
\ee
\be
e_3 = \frac{1}{\sqrt{(n^2 +m^2 t^2)(1+t^2)}} (n\partial_\xi - 
m\partial_\varphi )
\ee
and for the corresponding one-forms
\be
\omega_1 = dt \, ,\quad \omega_2 = t\sqrt{\frac{1+t^2}{n^2 +m^2 t^2}}
(m d\xi + nd\varphi )
\ee
\be
\omega_3 = \sqrt{\frac{1+t^2}{n^2 +m^2 t^2}} (nd\xi - mt^2 d\varphi ) .
\ee
The horizontal metric now is
\be
g_{\rm h}= dt \otimes dt + t^2 \frac{1+t^2 }{n^2 +m^2 t^2} (md\xi + nd\varphi )
\otimes (md\xi + nd\varphi )
\ee 
and is not yet manifestly conformally equivalent to the target space metric.
However, the horizontal metric only depends on the ``horizontal'' coordinates
$S=t$ and $\sigma =m\xi +n\varphi$ and, therefore, certainly {\em is}
conformally equivalent to the target space metric, because it is a
well-known fact that two different metrics on a two-dimensional surface
with a given topology are always conformally equivalent
(see, e.g., Theorem 13.1.1 in \cite{DFN}). All we have to do
is to find the coordinate transformation from the horizontal coordinates 
$(S,\sigma)$ to some new coordinates $(\tilde S , \tilde \sigma)$ such
that the conformal equivalence becomes manifest. This shows that the
initial problem must have a solution, i.e., higher Hopf maps, related
to the Hopf maps (\ref{chimn}), 
which solve the static eikonal equation, {\em must} exist. 

Explicitly, a transformation $(t,\sigma )\to (\tilde t (t), \sigma )$ is
sufficient such that
\be
g_{\rm h}= \frac{t^2}{\tilde t^2} \frac{1+t^2 }{n^2 +m^2 t^2}
[d\tilde t \otimes d\tilde t + \tilde t^2
(md\xi + nd\varphi )
\otimes (md\xi + nd\varphi ) ] .
\ee 
Therefore, $\tilde t$ has to obey 
\be \label{eq-tite}
(dt)^2 = \frac{(1+t^2)t^2}{(n^2 +m^2 t^2 )\tilde t^2} (d\tilde t)^2 \quad
\Rightarrow \quad \frac{1}{\tilde t} \frac{d\tilde t}{dt} =
\frac{1}{t}\sqrt{\frac{n^2 +m^2 t^2}{1+t^2}} .
\ee
Re-introducing the variable $\eta$ and using $\tilde t (t)=\tilde t(\sinh
\eta )\equiv f(\eta)$, the above equation (\ref{eq-tite}) leads
to equation (\ref{eq-efet}) with the solution (\ref{hi-sol}).

We want to close with two remarks. Firstly, the geometric setting 
developed above easily leads to more Hopf maps which solve the
static eikonal equation. Obviously, the sufficient
condition for the existence
of a solution related to a given Hopf map is that the induced horizontal
metric should be expressible - up to a local scale factor - in terms of the
``horizontal'' coordinates $(S,\sigma)$ (where the Hopf map is $\chi =
Se^{i\sigma}$). Once this condition is met, the solution can be found
by transforming to new ``horizontal'' coordinates $(\tilde S, 
\tilde \sigma )$
such that the horizontal metric is manifestly conformally equivalent 
to the target space metric. This transformation
is always possible for genuine Hopf maps. 
The simplest example of this type for the generation of new solutions is the 
composition
of existing solutions with maps $S^2 \to S^2$, i.e., the choice of new
complex-valued functions $ \chi' = F (\chi)$, where $F (\cdot )$ is
a holomorphic function (e.g. a rational map), and $\chi$ is a solution.
However, we already found these solutions from the symmetries of the
static eikonal equation in Section 2.

Secondly, we want to remark that the above Hopf maps (\ref{chimn}) do,
in fact, provide genuine Riemannian submersions from the three-sphere
$S^3$ to some two-dimensional target spaces with the topology of the
two-sphere but, in general,  metrics different from the two-sphere
(except for the simplest case $m=n=1$, which provides a Riemannian
submersion from $S^3$ to $S^2$, see, e.g., \cite{GLP,ES1}). 
This may
be understood from what we said above by noting that the local
scale factor
$q^{-2}$, which is present in the metric on $\RR^3$ (see (\ref{met3d})),
and which cannot be expressed in terms of the ``horizontal'' coordinates
alone, is absent for the metric on $S^3$.   

\vspace{0.5cm}

{\large\bf Acknowledgement:} The author thanks J. S\'anchez-Guillen for helpful
remarks. Further, support from 
MCyT (Spain), FEDER and Junta de Galicia (FPA2002-01161),
from the EC Commission grant HPRN-CT-2002-00325, as well as from the
Austrian START award project FWF-Y-137-TEC  
and from the  FWF project P161 05 NO 5 of N.J. Mauser is acknowledged.

\end{document}